\documentclass{article}
\usepackage{spconf,amsmath,graphicx}
\usepackage{booktabs}
\usepackage[shortlabels]{enumitem}
\usepackage{subcaption}
\usepackage{MnSymbol}
\usepackage{wasysym}
\DeclareMathOperator*{\argmax}{arg\,max}
\usepackage{dsfont}
\usepackage{hyperref,url,latexsym,color,soul}
\usepackage{setspace}
\usepackage{fancyhdr}
\makeatletter
\def\blfootnote{\gdef\@thefnmark{}\@footnotetext}
\makeatother

\title{BOFFIN TTS: Few-Shot Speaker Adaptation by Bayesian Optimization}%
\name{Henry B. Moss$^{\clubsuit }$\sthanks{This research was completed during an internship at Amazon Research. \hfill Correspondence to h.moss@lancaster.ac.uk and agvatsal@amazon.com.} \hspace{0.5em} Vatsal Aggarwal$^{\diamondsuit}$ \hspace{0.5em} Nishant Prateek$^{\diamondsuit}$ \hspace{0.5em} Javier Gonz\'alez$^{\diamondsuit}$ \hspace{0.5em} Roberto Barra-Chicote$^{\diamondsuit}$}

			\address{$^{\clubsuit}$ STOR-i Centre for Doctoral Training, Lancaster University \quad
			    $^{\diamondsuit}$Amazon Research Cambridge}
			    
\begin{document}
%
\maketitle
\begin{abstract}
We present BOFFIN TTS (Bayesian Optimization For FIne-tuning Neural Text To Speech), a novel approach for few-shot speaker adaptation. Here, the task is to fine-tune a pre-trained TTS model to mimic a new speaker using a small corpus of target utterances. We demonstrate that there does not exist a  \textit{one-size-fits-all} adaptation strategy, with convincing synthesis requiring a corpus-specific configuration of the hyper-parameters that control fine-tuning. By using Bayesian optimization to efficiently optimize these hyper-parameter values for a target speaker, we are able to perform adaptation with an average $30\%$ improvement in speaker similarity over standard techniques. Results indicate, across multiple corpora, that BOFFIN TTS can learn to synthesize new speakers using less than ten minutes of audio, achieving the same naturalness as produced for the speakers used to train the base model.
\end{abstract}
\begin{keywords}
text-to-speech, speaker adaptation, \\ \rightline{Bayesian optimization, transfer learning} 
\end{keywords}
\section{Introduction}
\label{sec:intro}
Given enough data, text to speech (TTS) systems can learn to convincingly mimic speakers across a wide range of acoustic and phonetic styles. However, training systems from scratch requires tens of hours of high-quality audio and reliable transcriptions, either from a single speaker to create speaker-specific models or spread across several speakers when training multi-speaker models \cite{latorre2019effect,oord2016wavenet,gibiansky2017deep,ping2017deep}. Training models on less data sacrifices quality and reliability \cite{chung2019semi}. 

To scale TTS catalogues across speakers for whom we have limited data, we adapt existing multi-speaker systems to generate new speakers - a well-studied form of transfer learning known as \textbf{speaker adaptation}\cite{article}. Adaptation is possible in scenarios where we have just minutes of target audio and partial phoneme coverage, as the robust representation of text and subsequent mappings to coherent speech are shared between the speakers \cite{latorre2019effect}. Only a small proportion of our network's capacity encodes speaker-specific information. We, therefore, need only enough utterances to learn \textbf{speaker identity} (the characteristics defining a target speaker's voice). 

Existing strategies for speaker adaptation fall into two broad categories. Many approaches use pre-trained auxiliary encoding networks to extract speaker characteristics to be combined with linguistic features as inputs to a TTS model \cite{li2017deep,taigman2017voiceloop,nachmani2018fitting,jia2018transfer}. In contrast, alternative approaches fine-tune the weights of existing multi-speaker models to synthesize new speakers \cite{arik2018neural,chen2018sample}. As fine-tuning provides the most natural adaptation across multiple TTS models \cite{arik2018neural,chen2018sample}, and is applicable to any existing system (without the need for training additional encoding networks), it is the focus of this report.

Our primary contribution is to demonstrate that successful speaker adaptation requires fine-tuning of adaptation hyper-parameters (henceforth referred to as the \textbf{adaptation strategy}) for each target speaker. We carefully tune the hyper-parameters governing adaptation and introduce two additional parameters not previously used for speaker adaption, demonstrating that the optimal hyper-parameter configuration depends subtly on the acoustic and phonetic properties of the target speaker alongside attributes of the target corpus (like audio-quality and size). For example, the amount of regularization required to prevent over-fitting (of which few-shot speaker adaptation is particularly susceptible \cite{chen2018sample}), depends on the quality and quantity of adaptation utterances. 

In this work, we formulate few-shot speaker-adaptation as an optimization problem - the task of finding appropriate hyper-parameter values for any given speaker. Our proposed BOFFIN\footnote{Boffin: British slang for a scientific expert.} TTS system automatically and efficiently solves this optimization problem through the hyper-parameter tuning framework of \textbf{Bayesian optimization} (BO), providing a fully automatic speaker-adaptation system suitable for general target speakers. BO has been shown to find high-performing hyper-parameters in competitively few model fits for many machine learning tasks \cite{snoek2012practical}, surpassing the performance of human tuners for problems in computer vision \cite{bergstra2013making}, natural language processing \cite{wang2015efficient}, and recently for reinforcement learning in AlphaGo \cite{chen2018bayesian}. However, BO has yet to see wide-spread use in TTS, where grid-based and random searches are still commonplace for hyper-parameter optimization. We hope that our successes with BO for speaker adaptation will encourage its more wide-spread use across TTS.

We evaluate BOFFIN TTS across three distinct scenarios, varying both the number of speakers in the base multi-speaker model and corpora audio-quality.

\section{System Description}
\vspace{-0.1in}
\subsection{Base Multi-Speaker Model}
\begin{figure}[ht]
	\begin{center}
		\centerline{\includegraphics[width=\columnwidth]{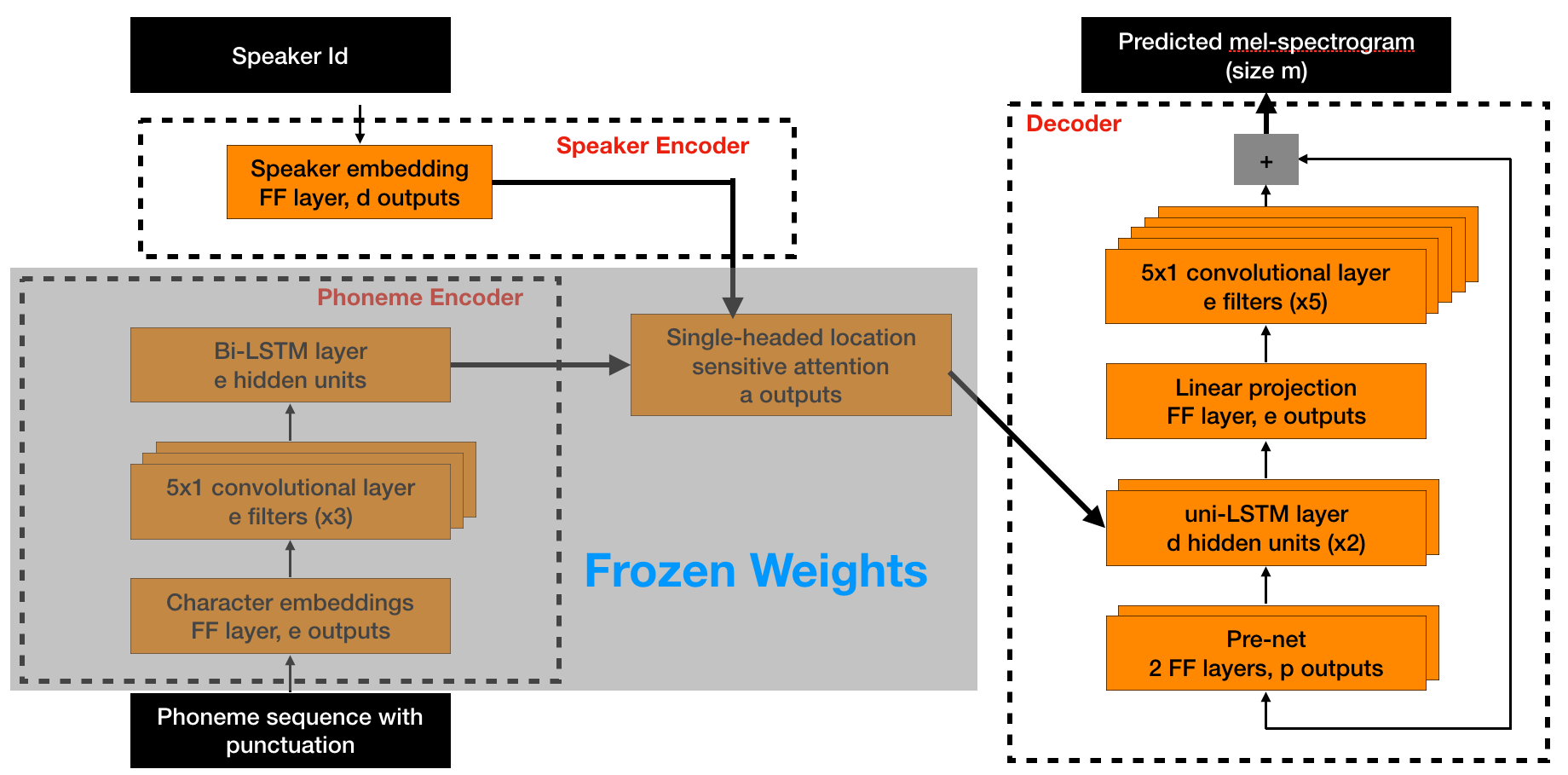}}
		\caption{Multi-speaker acoustic model architecture.}
		\label{architecture}
	\end{center}
	\vskip -0.6in
\end{figure}

\vspace{-0.05in}
\label{sec:system}
Our \textbf{base model} (the model we adapt to target speakers) is a Tacotron2 \cite{wang2017tacotron} style multi-speaker system explained in detail by \cite{latorre2019effect,prateek2019other}, consisting of an acoustic context-generation model and neural vocoder. Our acoustic model relies on an attention-based sequence-to-sequence network to generate context sequences (represented as mel-spectrograms) from input texts (see Figure \ref{architecture}).  Unlike Tacotron2 which models raw graphemes, we pre-process input text with a grapheme-to-phoneme module. To condition on individual speakers, we learn a speaker-embedding from a one-hot-encoding of speaker IDs (following \cite{oord2016wavenet}). This dense representation of speaker characteristics is presented to the attention module alongside encoded input text, to be decoded as a speaker-specific mel-spectrogram. Model weights are tuned with an ADAM optimizer to minimize the teacher-forced L1 loss between predicted and extracted mel-spectrograms. To complete the TTS pipeline, we convert mel-spectograms to waveforms using the multi-speaker neural vocoder of \cite{lorenzo2018robust}. This vocoder is trained across 74 speakers and suitable for generating natural speech for our wide-range of adaptation speakers. 
\vspace{-0.1in}

\subsection{Base-line Speaker Adaptation System}
\vspace{-0.05in}
Existing approaches for speaker adaptation by fine-tuning, although targeting different TTS architectures \cite{arik2018neural,chen2018sample}, all share the same approach which we apply to our chosen model to form a \textbf{base-line} adaptation system. To synthesize speakers not present in the training corpus, we continue the same learning process used to train the base model, but replace the training data with utterances of only the target speaker to allow the fine-tuning of weights and the learning of a new speaker embedding with respect to this new data. To avoid over-fitting to small collections of target utterances, we hold-out $20\%$ of adaptation data to form a validation set for early-stopping. From extensive human tuning, we know that the hyper-parameter configuration chosen for our base-model is capable of producing high-quality synthesis (achieving higher than four MOS naturalness scores for several speakers). We, therefore, expect this hyper-parameter configuration to form a competitive base-line for adaptation. Nevertheless, we later demonstrate that we can achieve a substantial improvement in adaptation quality using BOFFIN TTS.
\vspace{-0.1in}
\section{BOFFIN TTS}
\vspace{-0.1in}
\label{sec:BO}
There are two key difference between BOFFIN TTS and the base-line adaptation system. We allow the hyper-parameters controlling our adaptation to change to suit the target-speaker and, crucially, propose a framework for finding their optimal configuration in an efficient and automatic manner.
\vspace{-0.1in}
\begin{figure*}
	\centering
	\begin{subfigure}[b]{0.22\textwidth}
		\centering
		\includegraphics[width=\textwidth]{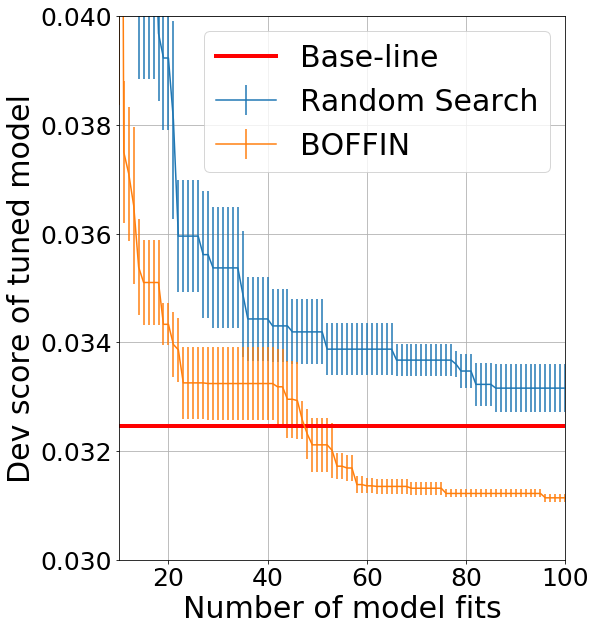}
		\caption{INTERNAL speaker A.}
	\end{subfigure}
	\begin{subfigure}[b]{0.22\textwidth}
		\centering
		\includegraphics[width=\textwidth]{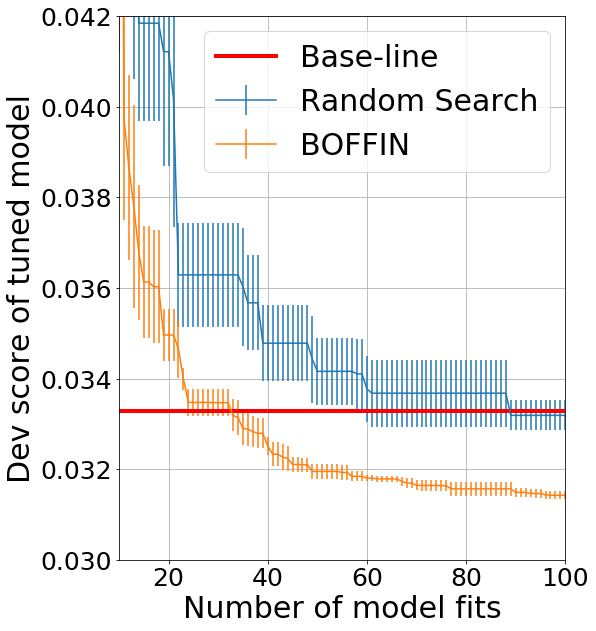}
		\caption{VCTK speaker p362.}
	\end{subfigure}
	\begin{subfigure}[b]{0.22\textwidth}
		\centering
		\includegraphics[width=\textwidth]{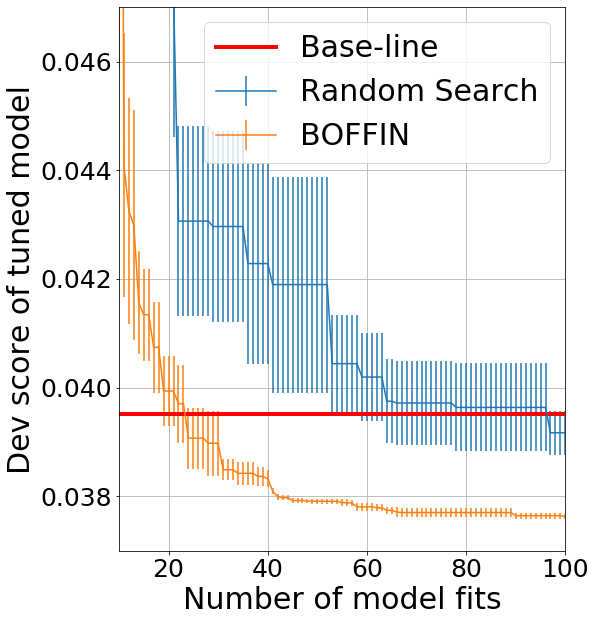}
		\caption{LibriTTS speaker 114.}
	\end{subfigure}
	\begin{subfigure}[b]{0.32\textwidth}
		\centering
		\includegraphics[width=\textwidth]{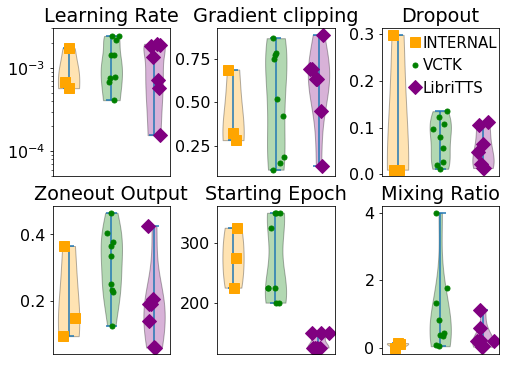}
		\caption{Tuned hyper-parameter values.}
		\label{BO_HP}
	\end{subfigure}	

	\caption{(a, b, c): Loss of the current best hyper-parameter configuration found by each system as we adapt to three randomly selected speakers from each corpora. We plot means and standard error for BOFFIN TTS and RS based on $5$ runs with different random seeds, alongside the loss achieved by the base-line adaptation system. (d): Hyper-parameter values chosen by BOFFIN TTS for multiple target speakers across three different data-sets. Each point represents a single speaker. We plot the six hyper-parameters whose optimal values show the largest variation across speakers.}\label{BO}
	\vskip -0.15in
\end{figure*}

\subsection{How Does BOFFIN TTS Control Adaptation?}
\vspace{-0.1in}
The key to effective adaptation is to learn characteristics of the target speaker without losing the generalizability of the base model (a phenomenon known as catastrophic forgetting). To this end, we believe there are nine key hyper-parameters that determine the success of adaptation.  These include seven parameters already widely used in machine learning to control learning dynamics (learning rate, batch size, decay-factor and gradient-clipping threshold) and to perform regularization (dropout and two zoneout parameters\cite{krueger2016zoneout}), alongside two parameters unique to BOFFIN TTS.

Although, tuning these seven standard hyper-parameters allowed us to learn the identity of the target speaker, the resulting models often show poor generalization capabilities. Therefore, we  propose two additional hyper-parameters. Firstly, we supplement our adaptation corpus, forming a tune-able ratio of target speakers to speakers already seen by the model (a simple approach to mitigate catastrophic forgetting known as a rehearsal method). Finally, we also tune which epoch of our trained base-model from which we begin adaptation. A base model before full convergence to the base speakers can provide a model more amenable for adaptation.

In addition to  hyper-parameter tuning, we also exploit the specific architecture of our chosen base-model. Rather than allowing our fine-tuning to update all model weights (as in \cite{chen2018sample}), we only allow fine-tuning of the weights in our speaker embedding and decoder modules (i.e those containing speaker-specific information, see Figure \ref{architecture}). We know that our encoder and attention modules are already able to facilitate synthesis across multiple speakers and we found that freezing their weights during adaptation led to more robust synthesis.

\vspace{-0.1in}
\subsection{How Does BOFFIN TTS Optimize Adaptation?}
\vspace{-0.1in}
Learning an optimal adaptation strategy for a target speaker is a difficult high-dimensional hyper-parameter optimization (HPO) problem. As is common in HPO, this optimization task is characterized by expensive evaluations (requiring a full adaptation to evaluate each single hyper-parameter configuration), a mixture of discrete and continuous variables, and a lack of analytical gradients for our objective function (the performance of adaptation) with respect to all our hyper-parameters. Consequently, we cannot apply gradient-based optimizers and the high-dimension of our turning task makes a simple grid-search computationally infeasible (and likely ineffective \cite{bergstra2012random}). We, therefore, use Bayesian optimization.

In a nutshell, BO is able to provide highly efficient HPO by using information from already evaluated hyper-parameter configurations to predict which untested configurations are `likely' to perform well and therefore should be next evaluated. In particular, to choose the ${t+1}^{th}$ hyper-parameter for evaluation, we fit a Gaussian process model \cite{rasmussen2003gaussian} to our $t$ collected configuration-evaluation pairs $\mathcal{D}_t=\{\textbf{x}_i,y_i\}_{i=1,..,t}$ across the hyper-parameter space $\mathcal{X}$, producing Gaussian predictions of performance at each configuration $\textbf{x}\in\mathcal{X}$ of $y(\textbf{x})|\mathcal{D}_t$. We then evaluate the configuration that we expect (according to our model) will provide the largest improvement over the best current best evaluation (with score $y'_t=\min_{i=1,..,t}y_i$), i.e we next evaluate configuration\begin{align}
\textbf{x}_{t+1}=&\argmax_{\textbf{x}\in\mathcal{X}}\mathds{E}_{y(\textbf{x})|\mathcal{D}_t}\left[\max(y_t'-y(\textbf{x}),0)|\mathcal{D}_t\right].
\label{EI}
\end{align}
For Gaussian processes, the inner expression of (\ref{EI}) and its gradients have convenient analytical forms (see \cite{eshahriari2015taking} for a comprehensive review of BO). Therefore, $\textbf{x}_{t+1}$ can be efficiently found using a standard gradient-based optimizer. 

We consider the performance of BOFFIN TTS when seeking to minimize L1 mel-spectogram loss across a held-out validation set of target speaker utterances. Although L1 loss does not necessarily correlate exactly with the perceptual quality of synthesized samples (as is the case for all objective TTS metrics), we found it informative enough to find hyper-parameters with high perceptual scores (Section \ref{sec:Results}). Adaptation to speakers from three different corpora is presented in Figure \ref{BO} (experimental details are discussed in Section \ref{sec:Results}). Our plots start after an initialization stage of $10$ random hyper-parameters, as this is required to provide a meaningful initial model across $\mathcal{X}$. Note that replacing BOFFIN TTS's BO component with random search (RS) fails to substantially improve upon our baseline (not speaker-specific) adaptation system. We need a sophisticated tuner like BO to find speaker-specific adaptation strategies. In addition, Figure \ref{BO_HP} shows that not only does the optimal hyper-parameter configuration vary between data-sets, but also across each individual speaker within each corpora. For example, our proposed \textit{Mixing Ratio} hyper-parameter requires larger values in general across the VCTK corpus than for our other corpora, however, the optimal  \textit{Mixing Ratio} still varies substantially across just the VCTK speakers.

\begin{table}
      \centering
	\begin{tabular}{l|lll}
		\textbf{System}            & \textbf{INTERNAL} & \textbf{VCTK} & \textbf{LibriTTS} \\ \hline
		base-synth     & 3.45 $\pm$ 0.08         &  3.76 $\pm$ 0.10   &     3.10   $\pm$ 0.10                  \\
		base-truth    & 3.84 $\pm$ 0.08         &     4.05 $\pm$ 0.08          &   4.10     $\pm$  0.08                \\
		adapt-synth  & 3.43 $\pm$ 0.10         &        3.6 $\pm$ 0.10            &   2.90   $\pm$   0.10                 \\
		adapt-truth & 4.05 $\pm$ 0.08         &       4.09 $\pm$    0.08        &    3.97  $\pm$    0.08               
	\end{tabular}
      	\caption{Comparing the mean naturalness scores achieved by BOFFIN TTS on target speakers (adapt-synth), by the base multi-speaker model on base speakers(base-synth), and by true audio for both target (adapt-truth) and base-model speakers (base-truth). We present each listener with samples across multiple base and adapted speakers and ask for a 5 point score from `completely unnatural' to `completely natural'. We print mean responses alongside $95\%$ confidence bounds.}
	\label{MOS}
	\vskip -0.15in
  \end{table}
\vspace{-0.2in}
\section{Results}
\vspace{-0.1in}
\label{sec:Results}
We have demonstrated that BOFFIN outperforms the base-line speaker adaptation system with respect to L1 loss. However, to investigate whether this lower score corresponds to an improvement in perceptual quality at inference time, we collected the perceptual evaluations of human listeners.
\vspace{-0.2in}
\subsection{Experimental Protocol}
\vspace{-0.1in}
To thoroughly test the performance of BOFFIN TTS, we consider three distinct corpora: (i) multi-speaker corpus with studio-quality recordings (referred to as INTERNAL\footnote{The internal corpus contains no customer voice recordings.}),  (ii) the open-source VCTK corpus \cite{veaux2017cstr}, and  (iii)  the LibriTTS audio-book corpus \cite{zen2019libritts}.  By considering a range of recording qualities and base-models with differing numbers of base speakers, we can understand the limitations of  using BOFFIN TTS in a variety of practical settings. The architecture of our base-model remains fixed except for the more challenging LibriTTS task, where we double the size of our speaker embedding to accommodate a larger collection of base speakers. BO is performed with the Python library Emukit \footnote{https://github.com/amzn/emukit}.

For each experiment, we adapt to $4$ unseen speakers (from the same corpora used to train the base-model) using a random sample of $100$ utterances (representing between $5$ and $10$ minutes of audio depending on the corpus), with $20\%$ retained as a validation set. To evaluate each system, we compare naturalness and achieved similarity to the target speaker using a MUltiple Stimuli with Hidden Reference and Anchor (MUSHRA) test \cite{recommendation2001method}. We also compare the naturalness achieved by BOFFIN TTS on target speakers with that achieved by the base multi-speaker model on its original speakers using a Mean Objective Score (MOS) test for naturalness. Each evaluation is presented to 25 native US listeners using Amazon Mechanical Turk. Statistical significance tests are performed at the p=0.01 level with Bonferroni-Holm corrections, using paired t and Mann-Whitney U tests for the MUSHRA and MOS evaluations respectively.
\vspace{-0.1in}
\subsection{Adaptation from a base-model with few speakers}
\vspace{-0.1in}
For our first experiment, we train a base-model on 4 male and 4 female proprietary speakers (each with 2.5k utterances) and adapt to 2 female and 2 child held-out speakers. Figure \ref{INTERNAL_sim} show that BOFFIN TTS is able to achieve  significant improvements in speaker similarity, with an improvement of $28\%$ over the base-line and $39\%$ over RS. Crucially, Figure \ref{INTERNAL_nat} shows BOFFIN TTS' improvement in similarity does not sacrifice perceptual quality, achieving a small but statistically significant improvement in naturalness over the base-line speaker adaptation system. Moreover, Table \ref{MOS} demonstrates that BOFFIN TTS is able to adapt to target speakers without a significant drop in perceptual quality from the base-model's speakers (learnt with $250$ times more data). 
\vspace{-0.1in}
\subsection{Adaptation from a moderately rich base-model }
\vspace{-0.1in}
We now consider a harder adaptation task; adapting to VCTK speakers with much higher variation in expressiveness, prosody and audio-quality than those in INTERNAL. Our base-model is trained on 22 speakers: 14 from VCTK  (with 400 utterances each) supplemented with the 8 already considered in our first experiment (added to provide a more robust base-model). We adapt to 4 unseen VCTK speakers. This challenging adaptation scenario necessitates target speaker-specific adaptation strategies, with BOFFIN TTS providing significant improvements of $57\%$ in similarity and $13\%$ in naturalness over the base-line (Figures \ref{VCTK_sim} and \ref{VCTK_nat}). Moreover, Table \ref{MOS} shows that BOFFIN TTS is once again able synthesize target speakers without a significant drop in naturalness than achieved for speakers already present in the base multi-speaker model.
\vspace{-0.1in}
\subsection{Adaptation from a  rich base-model}
\begin{figure}
	\centering
	\begin{subfigure}[b]{0.23\textwidth}
		\centering
		\includegraphics[width=\textwidth]{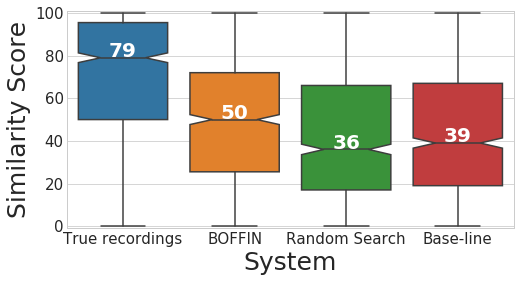}
		\caption{Similarity INTERNAL.}
		\label{INTERNAL_sim}
	\end{subfigure}
	\begin{subfigure}[b]{0.23\textwidth}
		\centering
		\includegraphics[width=\textwidth]{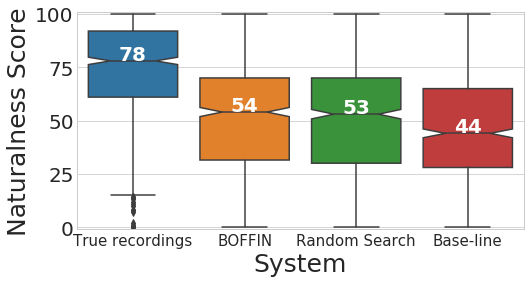}
		\caption{Naturalness INTERNAL.}
		\label{INTERNAL_nat}
	\end{subfigure}
	\begin{subfigure}[b]{0.23\textwidth}
		\centering
		\includegraphics[width=\textwidth]{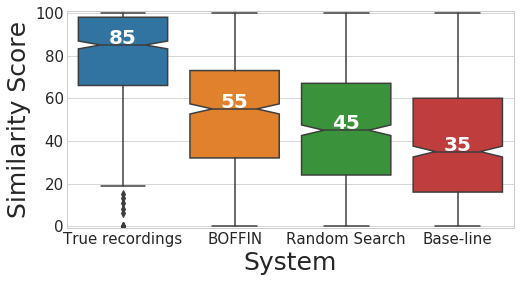}
		\caption{Similarity VCTK.}
		\label{VCTK_sim}
	\end{subfigure}
	\begin{subfigure}[b]{0.23\textwidth}
		\centering
		\includegraphics[width=\textwidth]{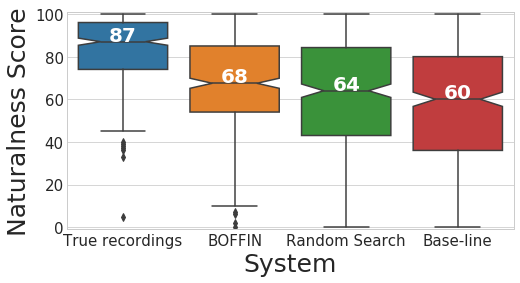}
		\caption{Naturalness VCTK.}
		\label{VCTK_nat}
	\end{subfigure}
	\begin{subfigure}[b]{0.23\textwidth}
		\centering
		\includegraphics[width=\textwidth]{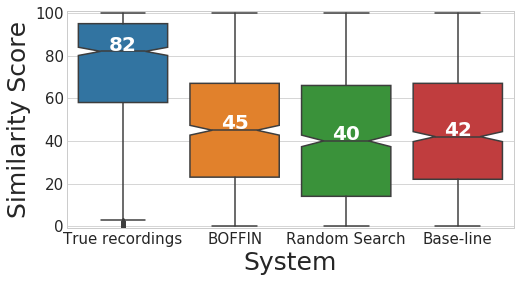}
		\caption{Similarity LibriTTS.}
		\label{LibriTTS_sim}
	\end{subfigure}
	\begin{subfigure}[b]{0.23\textwidth}
		\centering
		\includegraphics[width=\textwidth]{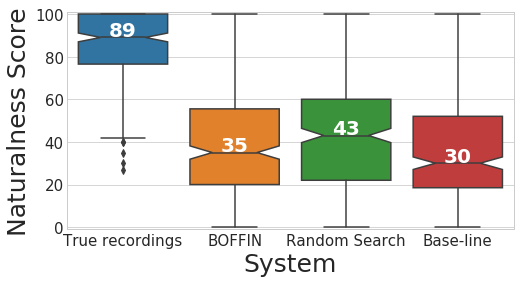}
		\caption{Naturalness LibriTTS.}
		\label{LibriTTS_nat}
	\end{subfigure}
	\caption{MUSHRA tests for speaker similarity and naturalness.  For similarity, we presented the same utterance synthesized by each system alongside a reference recording of the target speaker on another utterance and requested a rating of each system between `definitely a different person' (0) and `definitely the same person' (100). For naturalness, we repeat without a reference recording and instead asked for ratings between `completely unnatural' and `completely natural'}
	\label{discretepics}
	\vskip -0.15in
\end{figure}

\vspace{-0.1in}
To understand the limitations of BOFFIN TTS, our final experiment considers an even larger base-model containing 200 speakers (each with 200 utterances) from LibriTTS. We adapt to 4 additional unseen libriTTS speakers. LibriTTS is derived from audio-books and so contain noise, artifacts, and highly expressive voices. Consequently, although BOFFIN TTS was able to adapt to target speakers without a statistically significant drop in naturalness over the speakers used to train the base-system (Table \ref{MOS}) (as is consistent with our other experiments), our base-model itself was of much lower quality than our other base-models, making it difficult for our MUSHRA listeners to make a statistically significant preference in similarity across all three systems (Figures \ref{LibriTTS_sim} and \ref{LibriTTS_nat}).
\vspace{-0.1in}
\section{Conclusion}
\vspace{-0.1in}
We propose the few-shot speaker-adaptation framework of BOFFIN TTS. By learning adaptation strategies custom to each target speaker, BOFFIN TTS can achieve higher speaker similarity than using a \textit{one-size-fits-all} adaptation strategy, particularly when adapting to challenging target speakers from high-performance multi-speaker models.
\label{sec:Discussion}

\newpage


\bibliographystyle{IEEEbib}
\setstretch{0}
\bibliography{main}

\end{document}